\documentclass[twocolumn,aps,showpacs]{revtex4}
\usepackage{amssymb}
\usepackage{amsmath,bm}
\usepackage{graphicx}
\usepackage[normalem]{ulem}
\usepackage[dvips]{color}
\usepackage{lipsum}

\begin{document}
\title{\protect Medium effects on pion production in heavy ion collisions}
\author{Zhen Zhang\footnote{zhenzhang$@$comp.tamu.edu}}
\author{Che Ming Ko\footnote{ko$@$comp.tamu.edu}}
\affiliation{Cyclotron Institute and Department of Physics and Astronomy, Texas A$\&$M University, College Station, Texas 77843, USA}
\date{\today}

\begin{abstract}
Within the framework of the relativistic Vlasov-Uehling-Uhlenbeck transport model based on 
the relativistic nonlinear NL$\rho$ interaction, we study pion in-medium effects on the $\pi^-/\pi^+$ ratio in Au+Au collisions at the energy of $E/A=400~\mathrm{MeV}$. These effects include the isospin-dependent pion $s$-wave and $p$-wave potentials, which are taken from calculations based on the chiral perturbation theory and the $\Delta$-hole model, respectively. We find that the $\pi^-/\pi^+$ ratio in this collision is suppressed by the pion $s$-wave potential but enhanced by the $p$-wave potential, with a net effect of a significantly suppressed $\pi^-/\pi^+$ ratio. Including also the in-medium threshold effects on $\Delta$ resonance production and decay and using a nuclear symmetry energy with a slope parameter $L=59~\mathrm{MeV}$ by reducing the coupling of isovector-vector $\rho$ meson to nucleon, our result is in good agreement with measured $\pi^-/\pi^+$ ratio from the FOPI Collaboration. We further investigate the pion in-medium effects on the ratio of charged pions as a function of their kinetic energies.
\end{abstract}

\maketitle

\section{Introduction}

The density dependence of nuclear symmetry energy $E_{\mathrm{sym}}(\rho)$ has profound impacts on both nuclear physics and astrophysics~\cite{Lat04,Ste05,Bar05,LCK08}. While the nuclear symmetry energy below and around the saturation density $\rho_0=0.16~\mathrm{fm}^{-3}$ has been relatively well constrained by various experimental probes, such as the isospin diffusion, nuclear masses, neutron skin thickness, and giant dipole resonances~\cite{Hor01a,Fur02,Wan13,Dan14,Zha13,Bro13,Zha15,Tsa09,Tsa12,Lat14,Oer16,
BALi17}, its high-density behavior remains very uncertain. Heavy ion collisions with neutron-rich nuclei provide a unique approach to study the symmetry energy at supra-saturation densities. In particular, the $\pi^-/\pi^+$ ratio in heavy ion collisions at energies near the pion production threshold in a nucleon-nucleon collision is believed to be a promising probe of the symmetry energy at high densities~\cite{BALi02}, with a softer symmetry energy giving a larger $\pi^-/\pi^+$ ratio. During the last decade, many theoretical studies based on various transport models have been devoted to extract the high-density behavior of nuclear symmetry energy from the FOPI experimental data~\cite{FOPI07,FOPI10} on  the $\pi^{-}/\pi^{+}$ ratio~\cite{Xia09,Fen10,Xie13}. However, these studies have led to quite different conclusions, and there is still no qualitatively consensus on this issue. More in-depth understanding of pion production, such as the threshold effects~\cite{Fer05, Son15}, impact of energy conservation~\cite{Coz16} and pion in-medium effects~\cite{Xu10,Xu13, Hon14,Guo15, Fen15}, is therefore important. 

In Ref.~\cite{Son15}, based on the relativistic Vlasov-Uhling-Uhlenbeck (RVUU) model including explicitly the different isospin states of nucleons, $\Delta$ resonances, and pions,  the threshold effect, i.e., the change of pion production threshold in nuclear medium as a result of nucleon and $\Delta$ resonance potentials, on the $\pi^{-}/\pi^{+}$ ratio and the total pion yield has been studied. A considerable increase of the $\pi^{-}/\pi^{+}$ ratio is observed after including the threshold effect, and, in particular, the threshold effect reverses the effect due to the stiffness of nuclear symmetry energy on the $\pi^{-}/\pi^{+}$ ratio.  In this study, the pion in-medium effect is, however, not considered. Since studies based on a thermal model have indicated that the pion in-medium effect on the $\pi^-/\pi^+$ ratio is comparable to that due to the symmetry energy~\cite{Xu10,Xu13}, it is thus important to include this effect in the transport model to better understand the symmetry energy effect on pion production in heavy ion collisions.  In this work, we extend the isospin-dependent RVUU model of Ref.~\cite{Son15} to include the isospin-dependent pion $s$-wave and $p$-wave potentials in nuclear medium, which we take from calculations based on the chiral perturbation theory~\cite{Kai01} and the $\Delta$-hole model~\cite{Bro75, Fre81, CMKo89}, respectively. The effects of pion potentials on the $\pi^-/\pi^+$ ratio in central Au + Au collisions at $E/A=400~\mathrm{MeV}$ are then investigated, and the results are compared with the experimental data from the FOPI Collaboration~\cite{FOPI10}.

This paper is organized as follows. In Sec. II, we review the pion dispersion relation in nuclear medium due to its $s$-wave and $p$-wave interactions with nucleons. We then discuss in Sec. III the decay widths of $\Delta$ resonances that include the pion in-medium effect. The extended RVUU model, which includes both the pion in-medium effect and the threshold effect, are presented in Sec. IV.  Results on pion production in heavy ion collisions based on the extended RVUU model are shown and discussed in Sec. V. Finally, we give a summary in Sec. VI. 

\section{Pion dispersion relation in nuclear medium}

In nuclear medium, a pion acquires a self-energy from its  $s$-wave and $p$-wave interactions with nucleons. As a result, the pion dispersion relation in nuclear medium becomes 
\begin{equation}\label{dispersion}
\omega^2 = m_{\pi}^2+k^2 +\Pi_S+\Pi_P(\omega,\bm{k}),
\end{equation}
where $m_{\pi}=138~\mathrm{MeV}$ is the mass of pion, and $\Pi_S$ and $\Pi_P$ are the self-energies of the pion due to its respective $s$-wave and $p$-wave interactions with nucleons.

\subsection{$S$-wave  pion self-energy}

\begin{figure}[bpt]
\centering
\includegraphics[width=0.8\linewidth]{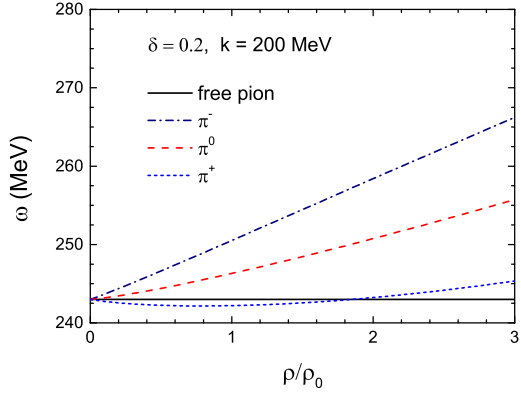}
\caption{(Color online) Density dependence of  the energy of a pion at fixed momentum $k=200~\mathrm{MeV}$ due to its $s$-wave interaction with nucleons in asymmetric nuclear matter of isospin asymmetry $\delta=0.2$. The black line indicates the case for a free pion. } \label{Fig:ES}
\end{figure}

For the pion $s$-wave self-energy in nuclear medium, it has been calculated  up to the two-loop approximation in chiral perturbation theory, and detailed expressions can be found in Ref.~\cite{Kai01}. In Fig.~\ref{Fig:ES}, we show the density dependence of pion energy after including the contribution from the pion $s$-wave self-energy. Here the pion momentum is fixed at 200~MeV and the isospin asymmetry of nuclear matter is taken to be $\delta=0.2$. It can be seen that for $\pi^{-}$ and $\pi^{0}$, the $s$-wave interactions are always repulsive, while for $\pi^{+}$ the $s$-wave interaction is attractive at low densities but repulsive at high densities. Since the pion $s$-wave self-energy is momentum independent, its contributions  can be absorbed into the mass term in Eq.(\ref{dispersion}) by defining the pion effective mass as $m_{\pi}^{\ast}= \sqrt{m_{\pi}^2+\Pi_S}$. We easily find $m_{\pi^-}^{\ast}>m_{\pi^0}^{\ast}>m_{\pi^+}^{\ast}$ in neutron rich matter.

\subsection{$P$-wave pion self-energy}

The pion $p$-wave self-energy in nuclear medium has usually been studied via the $\Delta$-hole model. For a pion in isospin state $m_t$ and of energy $\omega$ and momentum $\bm{k}$ in nuclear medium, its self-energy due to the $p$-wave interaction is given by~\cite{Bro75, Fre81,CMKo89}
\begin{eqnarray}
\Pi_0^{m_t} &\approx & \frac{4}{3}\left(\frac{f_{\Delta}}{m_{\pi}}\right)^2
\sum_{m_{\tau}}\left\vert \left\langle \frac{3}{2},m_t+m_{\tau}
\right\vert \left. 1, m_t; \frac{1}{2}, m_{\tau} \right\rangle\right\vert^2 \notag\\ 
&& \times \rho_{m_{\tau}}k^2 \frac{\omega_{\Delta}}{\omega_{m_{t}}^2-\omega_{\Delta}^2}e^{-2k^2/b^2}\notag \\
&=&\frac{8}{9}\left(\frac{f_{\Delta}}{m_{\pi}}\right)^2k^2\rho\left(1-\frac{m_t}{2}\delta\right)
\frac{\omega_{\Delta}}{\omega_{m_{t}}^2-\omega_{\Delta}^2}e^{-2k^2/b^2},\nonumber\\
\end{eqnarray}
with 
\begin{equation}
\label{Eq:EDH}
\omega_{\Delta} =\frac{k^2}{2m_{0}}+m_{0}-m_{N}.
\end{equation}
Here $m_{N}=939~\mathrm{MeV}$ and $m_{0}=1232~\mathrm{MeV}$ are the mass of nucleon and $\Delta$  resonance, respectively; $\rho=\rho_n+\rho_p$ is the nuclear density; $\delta=(\rho_n-\rho_p)/\rho$ is the isospin asymmetry; $f_{\Delta}\simeq2.3$ is the $\pi N \Delta$ coupling constant;  $b\simeq 7m_{\pi}$ is the cutoff of the $\pi N \Delta$ form factor; and $ \left\langle \frac{3}{2},m_t+m_{\tau}\right\vert \left. 1, m_t; \frac{1}{2}, m_{\tau} \right\rangle$ is the Clebsch-Gordan coefficient from the  isospin coupling of a pion with nucleon and $\Delta$ resonance. 

\begin{figure}[!bpt]
\centering
\includegraphics[width=1\linewidth]{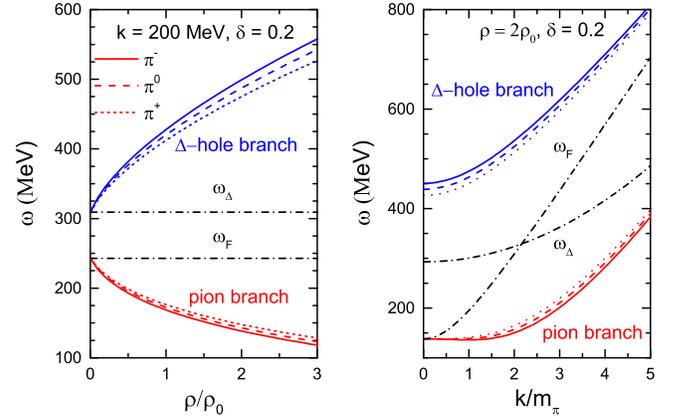}
\caption{(Color online) Density dependence of the energy of a pion at fixed momentum $k=200~\mathrm{MeV}$ (left window) and momentum dependence of its energy at fixed density $\rho=2\rho_0$ (right window) due to its $p$-wave potential in asymmetric nuclear matter of isospin asymmetry  $\delta=0.2$. The curves labeled by $\omega_{\mathrm{F}}$ and $\omega_{\Delta}$ are energies of free pion and $\Delta$-hole state given by Eq.~(\ref{Eq:EDH}), respectively.}
\label{Fig:EP}
\end{figure}

Including the short-range $\Delta$-hole interaction via the Migdal parameter $g^{\prime}\simeq0.6$~\cite{Bro75} leads to the modified pion self-energy, 
\begin{equation}
\Pi_P^{m_t} = \frac{\Pi_0^{m_t}}{1-g^{\prime}\Pi_0^{m_t}/k^2}.
\end{equation}
Because of the $p$-wave self-energy, the pion energy turns out to be 
\begin{equation}
\label{Eq:Ep}
\omega^2_{m_t} = \frac{1}{2}\left[
\omega_0^2+\hat{\omega}^2\pm\sqrt{(\omega_0^2-\hat{\omega}^2)^2+4k^2B\omega_{\Delta}}\right],
\end{equation}
with
\begin{eqnarray}
\omega_0&=&\sqrt{m_{\pi}^2+k^2},\label{Eq:E0} \\
\hat{\omega}&=&\sqrt{\omega_{\Delta}^2+g^{\prime}B\omega_{\Delta}}\label{Eq:Ehat},\\
B&=&\frac{8}{9}\left(\frac{f_{\Delta}}{m_{\pi}}\right)^2\rho\left(1-\frac{m_t}{2}\delta\right)e^{-2k^2/b^2}. 
\end{eqnarray}
From Eq.(\ref{Eq:Ep}), we see that there exist two branches of eigenstates, and the low(high)-energy branch is normally called the pion($\Delta$-hole) branch. Each branch consists of  a pion component and a $\Delta$-hole component with the probability for the pion component  given by
\begin{equation}
S =\frac{1}{1-\partial\Pi_P^{m_t}/\partial\omega^2}.
\end{equation}

We depict in Fig.~\ref{Fig:EP} the density dependence (left window) and momentum dependence (right window) of pion energies for the two branches in asymmetric nuclear matter of $\rho=2\rho_0$ and $\delta=0.2$. Compared with that of $\pi^{+}$, the dispersion relation of $\pi^{-}$ is  softer in the pion branch but stiffer in the $\Delta$-hole branch.

\subsection{Pion dispersion relation}

\begin{figure}[!bpt]
\includegraphics[width=1\linewidth]{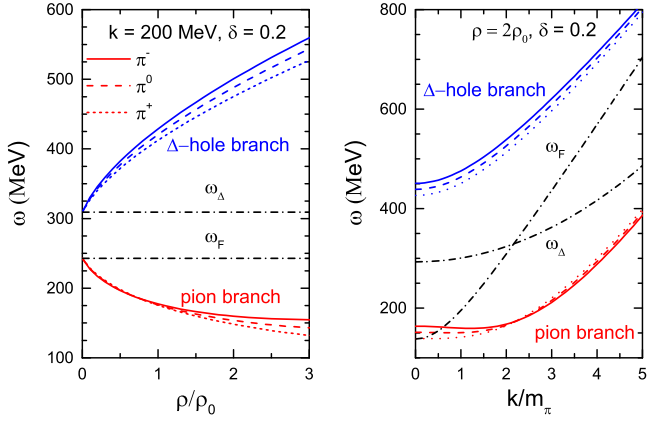}
\caption{(Color online) Same as Fig.~\ref{Fig:EP} but with the inclusion of both pion $s$-wave and $p$-wave potentials. }
\label{Fig:ESP}
\end{figure}

The pion dispersion relation in a nuclear medium including contributions of both $s$-wave and $p$-wave self-energies can be obtained from Eqs.~(\ref{Eq:Ep}) to ~(\ref{Eq:Ehat}) by substituting $m_{\pi}$ in Eq.(\ref{Eq:E0}) with $m_{m_t}^{\ast}$. We exhibit in Fig.~\ref{Fig:ESP} the density and momentum dependence of the resulting pion energy in asymmetric nuclear matter of $\rho=2\rho_0$ and $\delta=0.2$. In the left window, we fix the momentum at $k=200~\mathrm{MeV}$, while in the right window the nucleon density is fixed at $\rho=2\rho_0$. Due to the $s$-wave self-energy, the $\pi^{+}$ energy is lower than the $\pi^{-}$ energy for the $\Delta$-hole branch and also at momenta below about $2.5m_{\pi}$ for the pion branch.

\section{$\Delta$ resonance decay width in nuclear medium}
\label{Sec:Width}
 
The pion mean-field potentials also affect the  decay width of $\Delta$ resonance. In this work, we calculate the modified $\Delta$ resonance decay width following the method used in Ref.~\cite{Xio93}. For a free $\Delta$ resonance of mass $m_{\Delta}$ and in isospin state $m_T$ decaying into a nucleon and a pion with four momenta $(E_N,\bm{p}_N)$ and  $(\omega, \bm{k})$, respectively, its width can be calculated from 
\begin{widetext}
\begin{eqnarray}
\Gamma(m_{\Delta},\rho)& =&\frac{1}{2m_{\Delta}}
\sum_{m_{t}}
 \int 
\frac{d^3\bm{p}_N}{(2\pi)^32E_N}\frac{d^3\bm{k}d\omega}{(2\pi)^3}\delta(\omega^2-k^2-m_{\pi}^2-\Pi_S^{m_t}+\Pi_P^{m_t}(\omega,\bm{k}))\vert\mathcal{M}\vert^2 (2\pi)^4\delta^3(\bm{p}_N+\bm{k})\delta(E_N+\omega-m_{\Delta})\notag\\
&=&\sum_{m_t,i}\frac{1}{8\pi \sqrt{m_N^2+k_{m_t,i}^2}m_{\Delta}\omega_{m_t,i}}S(k_i,\rho)\vert \mathcal{M} \vert^2\left\vert \frac{\bm{k}_{m_t,i}}{\sqrt{m_N^2+k_{m_t,i}^2}}
+\frac{d\omega_{m_t,i}}{d\bm{k}_{m_t,i}}\right\vert^{-1}\label{Eq:Gamma}.
\end{eqnarray}
\end{widetext}
In the above, the summations $m_t$ and $i$ are over the isospin state of pion and its two branches of dispersion relation, respectively. The momentum of a pion in isospin state $m_t$ and in the $i$th branch is denoted by $\bm{k}_{m_t,i}$, and its magnitude can be determined by solving 
\begin{equation}
\label{Eq:PiP}
m_{\Delta}=\sqrt{m_N^2+\bm{k}^2}+\omega_{m_t,i}(\vert \bm{k}\vert,\rho).
\end{equation} 

In Eq.(\ref{Eq:Gamma}), the invariant matrix element for the decay of a $\Delta$ resonance in its rest frame is 
\begin{equation}
\vert\mathcal{M}\vert^2 =\frac{C}{4}\left( \frac{f_{ \Delta}}{m_{\pi}}\right)^2\text{Tr}[P^{\mu\nu}(p_{\Delta})
P_N(p_N)]k_{\mu}k_{\nu}e^{-2\bm{k}^2/b^2},
\end{equation}
where $C= \left\vert \left\langle \frac{3}{2},m_T\right\vert \left. 1, m_t; \frac{1}{2}, m_{T}-m_t \right\rangle\right\vert^2 $ is the square of the Clebsh-Gordan coefficient from the isospin coupling;  $p_{\Delta}$, $p_N$, and $k_{\mu}$ $p_{\mu}$ are the four-momenta of the $\Delta$ resonance, nucleon, and pion, respectively; and the factor 1/4 comes from the average over the  spin of $\Delta$ resonance. Using the well-known  projection operators for nucleon and $\Delta$ resonance,
\begin{equation}
P_N(p_N)= /\kern-0.5em p_N+m_N,
\end{equation}
and 
\begin{eqnarray}
P^{\alpha\beta}(p_{\Delta})& =&(/\kern-0.5em p_{\Delta}+m_{\Delta})
\left(g^{\alpha\beta}-\frac{2p_{\Delta}^{\alpha}p_{\Delta}^{\beta}}{3m_{\Delta}^2}\right.\notag  \\ 
& &\left. -\frac{\gamma^{\alpha}\gamma^{\beta}}{3}
+\frac{p_{\Delta}^{\alpha}\gamma^{\beta}-p_{\Delta}^{\beta}\gamma^{\alpha}}{3m_{\Delta}}\right),
\end{eqnarray}
we have
\begin{eqnarray}
\vert \mathcal{M}\vert^2=\frac{2C}{3}\left(\frac{f_{\Delta}}{m_{\pi}} \right)^2(m_N+E_N)m_{\Delta}\bm{k}^2 e^{-2\bm{k}^2/b^2}.
\end{eqnarray}
Taking into account the short-range $\Delta$-hole interaction modifies the invariant matrix element to
\begin{equation}
\label{Eq:Mat}
\vert \mathcal{M}\vert^2=\frac{2C}{3}\left(\frac{f_{\Delta}}{m_{\pi}} \right)^2(m_N+E_N)m_{\Delta}\left(
\frac{\bm{k}}{1-g^{\prime}\chi}\right)^2 e^{-2\bm{k}^2/b^2}.
\end{equation}

\begin{figure} [!hbt]
\centering 
 \includegraphics[width=0.9\columnwidth]{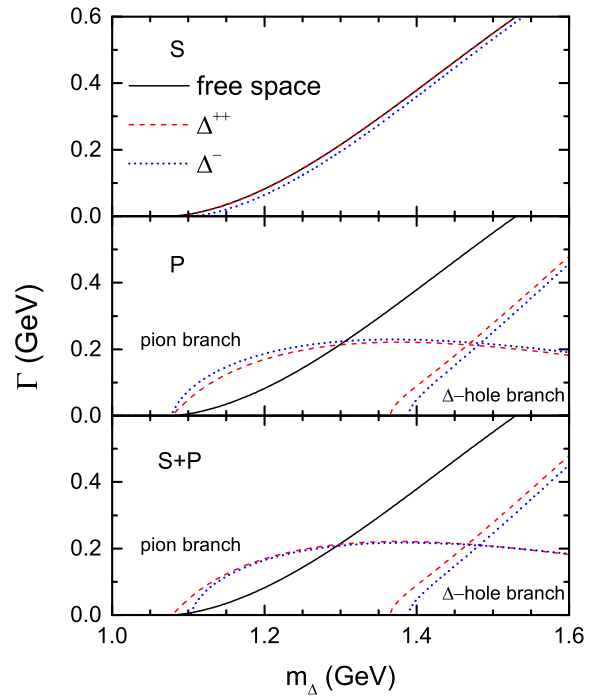} 
\caption{(Color online) Decay widths of $\Delta^{++}$ and $\Delta^-$ as functions of their masses after the inclusion of the pion  $s$-wave potential (S), the $p$-wave potential (P), and both the $s$-wave and $p$-wave potentials (S+P) in asymmetric nuclear matter of $\rho=2\rho_0$ and $\delta= 0.2$. The case of a $\Delta$ resonance in free space is shown as  black solid lines. Here nucleons and $\Delta$ resonances are treated as free particles.} \label{Fig:DltWid}
\end{figure}

In free space, we obtain from Eqs.~(\ref{Eq:Mat}) and (\ref{Fig:DltWid}) the  decay width of $\Delta$ resonance,
\begin{equation}
\Gamma_0=\frac{k^3}{12\pi}\left(\frac{f_{\Delta}}{m_{\pi}}\right)^2\frac{m_N+E_N}{m_{\Delta}}e^{-2\bm{k}^2/b^2}.
\end{equation}
For a $\Delta$ resonance of mass $m_{0}=1.232~\mathrm{GeV}$, the above equation gives $\Gamma_0=0.12~\mathrm{GeV}$, which is the same as the empirical value.

In Fig.~\ref{Fig:DltWid}, we show the decay widths of $\Delta^{++}$ and $\Delta^-$ as functions of their masses after the inclusion of the pion $s$-wave potential (S), the $p$-wave potential (P), and both the $s$-wave and $p$-wave potentials (S+P) in asymmetric nuclear matter of $\rho=2\rho_0$ and $\delta= 0.2$. Here nucleons and $\Delta$ resonances are treated as free particles. For comparison, the case of $\Delta$ resonance in free space is shown as  black solid lines.

It is seen that including the effect of  pion $s$-wave potential increases the $\Delta^-$ decay threshold and reduces its decay width. This is because the $s$-wave potential increases the effective mass of $\pi^-$ and therefore reduces the $\pi^-$ momentum $k$. The $\Delta^+$ width remains, on the other hand, almost unchanged since the pion $s$-wave potential only has little effects on the $\pi^+$ effective mass around $2\rho_0$. Therefore, including the pion $s$-wave potential is expected to suppress the $\pi^{-}/\pi^{+}$ ratio in heavy ion collisions.

From the middle panel of Fig.~\ref{Fig:DltWid}, we see that the threshold for  $\Delta$ resonance to decay into a pion in the $\Delta$-hole branch is very large ($\sim1.36~\mathrm{GeV}$), making its contribution less important than that due to decay to the pion branch.  Therefore, in the present work that focuses on low energy heavy ion collisions, we neglect the decay of  $\Delta$ resonance into a pion in the $\Delta$-hole branch. For the decay of  $\Delta$ resonance to a pion in the pion branch, we find that its width is increased by the pion $p$-wave potential  for small $\Delta$ resonance masses but reduced for large $\Delta$ resonance masses. This is due to the competition between the decreasing probability factor $S$ and increasing decay matrix element and phase space with increasing $\Delta$ resonance mass. For low energy heavy ion collisions,  the process $\Delta\rightarrow N+\pi$ is therefore enhanced by the pion $p$-wave potential.  In particular, since the pion dispersion relation in nuclear medium is softened by the pion-nucleon $p$-wave interaction, the decay width of $\Delta$ resonance becomes larger and thus enhances its decay. Although the pion $s$-wave potential generally decreases the width of a $\Delta$ resonance, including both pion $s$-wave and $p$-wave potentials still leads to an enhanced decay of $\Delta$ resonances as shown in the bottom panel of Fig.~\ref{Fig:DltWid}. 

In the present study, we also include the $\Delta$ resonance and nucleon in-medium effects by substituting their masses $m_n$ and $m_{\Delta}$ in Eq.~(\ref{Eq:Mat}) with corresponding effective masses $m_N^{\ast}$ and $m_{\Delta}^{\ast}$ given in Ref.~\cite{Son15}. In this case, instead of Eq.~(\ref{Eq:PiP}), the  momentum of the pion from  the decay of a $\Delta$ resonance in nuclear medium  is determined by solving 
\begin{equation}
\label{Eq:DNPi}
m_{\Delta}+\Sigma^0_{\Delta} = \sqrt{{m_N^{\ast}}^2+k^2}
+\Sigma^0_{N} +\omega(k,\rho),
\end{equation}
where $\Sigma_N^0$ and $\Sigma_{\Delta}^0$ are the time components of nucleon and $\Delta$ resonance vector self-energies in the laboratory frame, respectively, which are also given in Ref.~\cite{Son15}. In the above, we have neglected the spatial components $\bm{\Sigma}_{\Delta}$ and $\bm{\Sigma}_{N}$ of the nucleon and $\Delta$ resonance vector self-energies since they are small compared with corresponding time components.

\section{Relativistic Vlasov-Uehiling-Uhlenbeck model}

To study the pion in-medium effects in heavy ion collisions, we extend the RVUU model~\cite{Son15} by including  them on the production and absorption of $\Delta$ resonances and pions as well as the propagation of pions.

\begin{figure} [!hbt]
\centering 
 \includegraphics[width=0.9\columnwidth]{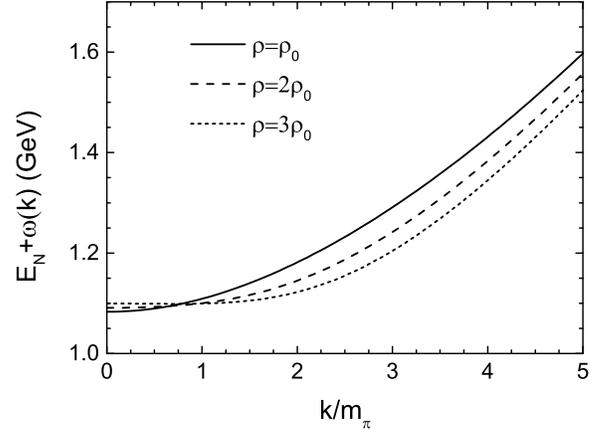} 
\caption{Sum of nucleon and pion energies $E_N+\omega(k)$ as a function of pion momentum $k$ in symmetric nuclear matter at different densities with the inclusion of both pion $s$-wave and $p$-wave potentials. Here the nucleon is treated as a free particle.} \label{Fig:MK}
\end{figure}

For the reaction $N+N^{\prime}\rightarrow N^{\prime\prime}+\Delta$, the pion mean-field potential affects its threshold via the minimum mass $m_{\Delta,\mathrm{min}}^{\ast}$ of $\Delta$ resonance that can decay into nucleon and pion, which can be determined according  to 
\begin{equation}
\label{Eq:MinDltMas}
m_{{\Delta,}\mathrm{min}}^{\ast} = \mathrm{min}_k\left\lbrace\sqrt{{m_N^{\ast}}^2+k^2}
+\Sigma^0_{N}-\Sigma^0_{\Delta} +\omega(k,\rho)\right\rbrace,
\end{equation} 
where the spatial components of nucleon and delta vector self-energies are again neglected. In free space, one has $\omega =\sqrt{m_{\pi}^2+k^2}$, and $m_{\Delta,\mathrm{min}}^{\ast}$ is just equal to $m_N+m_{\pi}$. While the effect of the pion $s$-wave potential can be easily included by replacing $m_{\pi}$ with $m_{\pi}^{\ast}$, it is nontrivial to include the effect of the pion $p$-wave potential. However, for low-energy heavy ion collisions, the effect of the pion $p$-wave potential on the minimum mass of $\Delta$ resonance can be safely neglected. To illustrate this point, we show in Fig.~\ref{Fig:MK} the momentum dependence of $\sqrt{m_N^2+k^2}+\omega(k)$ with the inclusion of both pion $s$-wave and $p$-wave potentials in symmetric nuclear matter at different densities. It is see that even at the high density of $\rho=3\rho_0$, the minimum mass is only slightly less ($<1$ MeV) than the value at $k=0$, i.e., $m_N+m_{\pi}^{\ast}$. Therefore, we neglect in this study the effect of the pion $p$-wave interaction  on the reaction $N+N^{\prime}\rightarrow N^{\prime\prime}+\Delta$, and evaluate the minimum mass of $\Delta$ according to
\begin{equation}
m_{\Delta,\mathrm{min}}^{\ast} =m_N^{\ast}+m_{\pi}^{\ast}
+\Sigma^0_{N}-\Sigma^0_{\Delta}. 
\end{equation} 
For $\Delta^{0}$ and $\Delta^{+}$, which may have different minimum masses in their two different 
decay channels, we use the smaller one to determine the production threshold of $
\Delta$ resonance.  Since $m_{\pi^-}^{\ast}>m_{\pi^+}^{\ast}$ in neutron-rich matter, the above 
equation thus indicates that the pion $s$-wave potential suppresses the 
production of $\Delta^-$ more than $\Delta^+$ and therefore may reduce the charged pion ratio 
in heavy ion collisions.

In the RVUU model, the $\Delta$ resonance is treated as a particle of various masses  according to the Lorentzian function~\cite{Dan91,Li93,LK95}. In the present study, this is generalized to
\begin{equation}
P(m_\Delta^{\ast})=\frac{p_f^{\ast}m_\Delta^{\ast}\Gamma_{\mathrm{tot}}(m_{\Delta}^{\ast})}{(m_{\Delta}^{\ast2}-m_{0}^{\ast2})^2+m_0^{\ast2}\Gamma_{\mathrm{tot}}^2(m_{\Delta}^{\ast})},
\end{equation}
where $m_{\Delta}^{\ast}$ and $m_0^{\ast}$ are the shifted $\Delta$ resonance mass and pole mass by the scalar mean field, $p_f^{\ast}$ is the kinetic momentum of the $\Delta$ resonance in the center of mass frame of final nucleon and $\Delta$ resonance, which is defined by their kinetic momenta, i.e., $\bm{p}_{N^{\prime\prime}}^{\ast}+\bm{p}_{\Delta}^{\ast}=0$, and the decay width $\Gamma_{\mathrm{tot}}$ is the total decay width calculated using the method introduced in Sec.\ref{Sec:Width}.

The RVUU model also includes the inverse reaction $N^{\prime\prime}+\Delta \rightarrow N+N^{\prime}$. Its  cross section is related to that of the reaction $N+N^{\prime}\rightarrow N^{\prime\prime}+\Delta$ via the detailed balance relation~\cite{Dan91,Li93,LK95},
\begin{eqnarray}
\sigma(N^{\prime\prime}\Delta\rightarrow N N^{\prime}) &=&\frac{m^{\ast}}{8 m_0^2}\frac{1}{1+\delta_{NN^{\prime}}}
\frac{p_i^{\ast 2}}{p_f^{\ast}}\sigma(N N^{\prime} \rightarrow N^{\prime\prime}\Delta)\notag \\
& &\times \left[ \int_{m_{\mathrm{min}}^{\ast}}^{m_{\mathrm{max}}^{\ast}}\frac{dm}{2\pi}P(m)\right]^{-1},
\end{eqnarray}
where $p_i^{\ast }$ is the nucleon kinetic momentum in the frame of $\bm{p}_{N}^{\ast}+\bm{p}_{N^{\prime}}^{\ast}=0$.

The maximum effective mass of $\Delta$, $m_{\Delta,\mathrm{max}}^{\ast}$, in above equation can be determined by evaluating the square of the center of mass energy 
\begin{eqnarray}
s&=&\left(\sqrt{m_{N^{\prime\prime}}^{*2}+{\bf p}_{N^{\prime\prime}}^{*2}}+\Sigma_{N^{\prime\prime}}^0+\sqrt{m_\Delta^{*2}+{\bf p}_\Delta^{*2}}+\Sigma_\Delta^0\right)^2\notag\\
&-&({\bf p}_{N^{\prime\prime}}+{\bf p}_\Delta)^2
\end{eqnarray}
in the frame of $\bm{p}_{N}^{\ast}+\bm{p}_{N^{\prime}}^{\ast}=\bm{p}_{N^{\prime\prime}}^{\ast}=\bm{p}_{\Delta}^{\ast} = 0$.  Denoting the nucleon and $\Delta$ resonance self-energies $\Sigma_N$ and $\Sigma_\Delta$ in this frame by $\Sigma_N^\prime$ and $\Sigma_\Delta^\prime$, respectively, we then have  
\begin{eqnarray}
s&=&(m_{N^{\prime\prime}}^{*2}+\Sigma_{N^{\prime\prime}}^{0\prime}+m_\Delta^{*2}+\Sigma_\Delta^{0\prime})^2-({\bm\Sigma}_{N^{\prime\prime}}^\prime+{\bm\Sigma}_\Delta^\prime)^2,\notag\\
\end{eqnarray}
which leads to
\begin{equation}\label{max}
m_{{\Delta,}\mathrm{max}}^{\ast}=\sqrt{s+\left(\bm{\Sigma}_{N^{\prime\prime}}^{\prime}+\bm{\Sigma}_{\Delta}^{\prime} \right)^2}-m_{N^{\prime\prime}}^{\ast}-\Sigma ^{0\prime}_{N^{\prime\prime}}-\Sigma^{0\prime}_{\Delta}.
\end{equation}

The decay of a $\Delta$ resonance during each time step of $dt$ in the RVUU model is treated by the Monte Carlo method with the decay probability given by
\begin{equation}
P=1-\mathrm{exp}[-dt\Gamma_{\mathrm{tot}}/\gamma],
\end{equation} 
where $\gamma= E_{\Delta}/\sqrt{E_{\Delta}^2-\bm{p}_{\Delta}^2}$ is the Lorentz factor for the $\Delta$ resonance. The charge state of emitted pion is chosen according to the branch ratio $\Gamma_{m_t}/\Gamma_\mathrm{tot}$ with $\Gamma_{m_t}$ being the partial decay width.  In determining the four-momentum of the pion, we assume that the pion is emitted isotropically in the frame $F^{\prime}$ of $\bm{p}_N^{\ast}+\bm{p}_{\pi}=0$. This frame can be obtained from the laboratory frame by a Lorentz transformation with the velocity
\begin{equation}
\beta = \frac{\bm{p}_N^{\ast}+{\bm{k}}}{E_N^{\ast}+\omega}
=\frac{\bm{p}_{\Delta}-\bm{\Sigma}_N}{E_{\Delta}-\Sigma_N^0},
\end{equation} 
where $\Sigma_N$ is the nucleon mean field in the laboratory frame. Here we use the conditions of canonical energy and momentum conservation, namely, $E_{\mathrm{\Delta}}=E_N+\omega$
and  $\bm{p}_{\Delta}=\bm{p}_N+\bm{k}$. After randomly generating a direction for the kinetic momentum $\bm{p}^{\ast \prime}_N$ of the emitted nucleon in the frame $F^{\prime}$, its kinetic momentum $\bm{p}^{\ast}_N$ in the laboratory frame can be determined if its magnitude  $p_N^{\ast \prime}$ is known. In terms of the pion three momentum,  $\bm{p}_{\Delta}-\bm{p}^{\ast}_N-\bm{\Sigma}_N$, $p_N^{\ast \prime}$ can then be determined from the energy conservation condition, 
\begin{equation}
E_{\mathrm{\Delta}}=\sqrt{m_N^{\ast 2}+\bm{p}^{\ast 2}_N}+
\Sigma_N+\omega(\rho,\vert\bm{p}_{\Delta}-\bm{p}^{\ast}_N-\bm{\Sigma}_N\vert).
\end{equation}

The motion of a pion in the RVUU model obeys the classical equations of motion, i.e.,
\begin{eqnarray}
\dot{\bm{r}}&=&\frac{d\omega}{d\bm{k}}, \\
\dot{\bm{k}}&=&-\nabla\omega.
\end{eqnarray}
Since $d\omega/dk$ is less than that of a free pion in both the pion and $\Delta$-hole branches, pions moves slower in nuclear medium .

\begin{figure}
\centering 
\includegraphics[width=1\columnwidth]{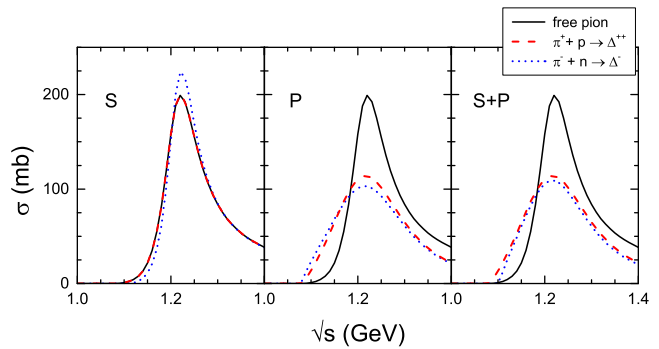} 
\caption{ (Color online) Cross sections of the reaction $\pi^++p\rightarrow \Delta^{++}$ and $\pi^-+n\rightarrow \Delta^{-}$ as functions of $\sqrt{s}$ with inclusion of the pion $s$-wave  potential (S), the $p$-wave potential (P) and both $s$-wave and $p$-wave potentials (S+P) in asymmetric nuclear matter of $\rho=2\rho_0$ and $\delta = 0.2$. For comparison, we also show the cross section for free pions as  black solid lines.  Here nucleons and $\Delta$ resonances are treated as if they are in free space.}\label{Fig:Snpid}
\end{figure}

A pion can be absorbed by a nucleon to form a  $\Delta$ resonance. In the case that the nucleon and $\Delta$ resonance are free particles, the cross section for the reaction $N+ \pi\rightarrow \Delta$ in the center of mass frame 
can be evaluated as
\begin{eqnarray}
\sigma & =& 
\int \frac{dm_{\Delta}}{2\pi}\mathcal{A}(m_{\Delta}) 
\int \frac{d^3\bm{p}_{\Delta}}{(2\pi)^3 2E_{\mathrm{\Delta}}} \frac{ \vert\mathcal{M}_{N\pi\rightarrow \Delta}\vert^2}{4(E_N\omega)\vert v_N-v_{\pi}\vert}\notag \\
& &\times (2\pi)^4
\delta^3(\bm{p}_N+\bm{k}-\bm{p}_{\Delta})\delta(E_N+\omega-E_{\Delta})\notag\\
&=& \frac{1}{8m_{\Delta}^2k}\vert\mathcal{M}_{N\pi\rightarrow \Delta}\vert^2\mathcal{A}(m_{\Delta}),
\end{eqnarray}
where  $\mathcal{A}(m_{\Delta})$ is the spectral function of $\Delta$ resonance~\cite{Dan91,Li93,LK95},
\begin{equation}
\label{Eq:SpDlt}
\mathcal{A}=\frac{4m_0^2\Gamma_{\mathrm{tot}}}{(m_{\Delta}^2-m_0^2)^2+m_0^2\Gamma_{\mathrm{tot}}^2}.
\end{equation}
In obtaining the second expression in the above equation, we have used the relation $\vert v_N-v_{\pi}\vert= \vert \bm{p_N}/E_N-d\omega/d{\bm{k}}\vert$ for the relative velocity between the nucleon and pion.
Using the detailed balance relation $\vert\mathcal{M}_{N\pi\rightarrow \Delta}\vert^2= 2\vert\mathcal{M}_{\Delta\rightarrow N\pi }\vert^2$ and Eq.~(\ref{Eq:Gamma}) results in the following expression for the pion absorption cross section by a nucleon:
\begin{eqnarray}
\sigma &=& \frac{8\pi}{k^2} \frac{1}{ S(k,\rho)}  \frac{m_0^2\Gamma\Gamma_{\mathrm{tot}}}{(m_{\Delta}^2-m_0^2)^2+m_0^2\Gamma_{\mathrm{tot}}^2}.
\end{eqnarray}

Medium effects due to the pion and nucleon potentials on the pion absorption cross section can be included in the above formula with the pion momentum $k$ calculated according to Eq.~(\ref{Eq:PiP}) and using the nucleon and $\Delta$ resonance effective masses. Taking the reactions $\pi^++p\rightarrow \Delta^{++}$  and  $\pi^-+n\rightarrow \Delta^{-}$ as examples and ignoring the nucleon and $\Delta$ mean-field potentials, we show in Fig.~\ref{Fig:Snpid} these cross sections as functions of the center of mass energy $\sqrt{s}$ of the scattering pion and nucleon, i.e.,  the mass $m_{\Delta}$ of produced $\Delta$ resonance, for the cases of including the pion $s$-wave potential (S), the $p$-wave potential (P), and both $s$-wave and $p$-wave potentials (S+P) in asymmetric nuclear matter of $\rho=2\rho_0$ and $\delta =0.2$. For comparison, the cross sections for free pions are shown as black solid lines. It can be seen that while the pion $p$-wave potential substantially reduces the pion absorption cross section, the effect of the pion $s$-wave potential is rather small, except around the  pole mass $1.232~\mathrm{GeV}$ of $\Delta$ resonance, where it leads to a considerable increase of the cross section for $\pi^-+n\rightarrow \Delta^{-}$.

\section{Results and discussions}

We study in this section the pion in-medium effects on the charged pion ratio in Au+Au collision at $E/A=400~\mathrm{MeV}$ in the RVUU model with the nucleon mean-field potentials based on the relativistic NL$\rho$ model~\cite{Liu02} and compare the results with experimental data from the FOPI Collaboration~\cite{FOPI10}. Since the charged pion ratio also depends on the stiffness of nuclear symmetry energy~\cite{BALi02} and is affected by the threshold effect~\cite{Son15}, we compare in the following results from  six different cases, i.e., (i) without the threshold and pion in-medium effects (free), namely, nucleons, $\Delta$ resonances and pions are treated as free particles in all reactions; (ii) with only the threshold effect (Th); (iii) with the threshold effect and the pion $s$-wave potential (S); (iv) with the threshold effect  and the pion $p$-wave potential (Th+P); (v) with the threshold effect and both the pion $s$-wave and $p$-wave potentials (Th+S+P); (vi) same as case (v) but with the coupling constant $f_{\rho}=0.95~\mathrm{fm}^2$ of the isovector-vector $\rho$ meson to nucleon in the NL$\rho$ model reduced to $f_{\rho}=0.43~\mathrm{fm}^2$. The latter reduces the slope parameter $L$ of the nuclear symmetry energy from $84~\mathrm{MeV}$ to $59~\mathrm{MeV}$, which is consistent with the average value from  analyses using various observables and methods~\cite{Oer16,BALi17}.

\begin{figure} [!hbt]
\centering 
 \includegraphics[width=0.8\columnwidth]{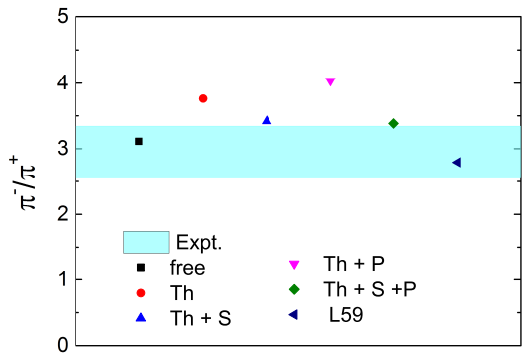} 
\caption{ (Color online) The $\pi^{-}/\pi^{+}$ ratio in Au+Au collisions at impact parameter of 1.4 fm and energy of $E/A=400~\mathrm{MeV}$ from the NL$\rho$ model in different cases (see text for details).  Experimental data~\cite{FOPI10} from the FOPI collaboration are shown as the cyan band.
} \label{Fig:Ratio}
\end{figure}
  
We include the threshold effect by following Ref.~\cite{Son15}.  As in Ref.~\cite{Son15}, we use a medium-dependent cross section for $\Delta$ production from the reaction $N+N\to N+\Delta$ by assuming the following density dependence:
\begin{equation}
\sigma_{NN\rightarrow\Delta N}(\rho_N)=\sigma_{NN\rightarrow \Delta N}(0) \mathrm{exp}(-A\sqrt{\rho_N/\rho_0}),
\end{equation}
where $\rho_N$ is the nucleon density and $A$ is an adjustable parameter. The cross section for $N+\Delta\rightarrow  N+N$ is also modified according to the detailed balance relation. We find that our model can well reproduce the experimental data on the total  pion yield in Au + Au collisions at $E/A=400~\mathrm{MeV}$ if the values $A=-2.35,~1.5,~1.4,~1.65 $, $1.5$ and $1.5$ are used  for the above six cases, respectively. The variation of the values for $A$ is due to the fact that the total pion yield is enhanced by the threshold effect and the pion $p$-wave potential, and suppressed by  the $s$-wave potential.  Because of the cancellation between the pion $s$-wave and $p$-wave potentials, and the fact that the $f_{\rho}$ only influences the isovector properties of nuclear matter, including both the pion $s$-wave and $p$-wave potentials and the reduction of $f_{\rho}$ does not significantly affect the total pion yield.

We show in Fig.~\ref{Fig:Ratio} the $\pi^-/\pi^+$ ratio in Au+Au collisions at impact parameter of 1.4 fm and energy of $E/A=400~\mathrm{MeV}$ from the above six different cases. It is seen that the threshold effect substantially increases the $\pi^-/\pi^+$ ratio by about $20\%$.  For the effects of pion potentials, we find that the pion $s$-wave potential reduces and the $p$-wave potential enhances the $\pi^-/\pi^+$ ratio. As a result, including both potentials leads to a significant decrease ($\sim 10\%$) of the $\pi^-/\pi^+$ ratio. The effect of $s$-wave potential can be easily understood since at densities below about $2\rho_0$, the effective masses of $\pi^-$ and $\pi^0$ increase while the $\pi^+$ effective mass slightly decreases.  The enhancement of the $\pi^-/\pi^+$ ratio after the inclusion of the pion $p$-wave potential is due to the softer dispersion relation of $\pi^{-}$ in the pion branch. The pion in-medium effects on the $\pi^{-}/\pi^{+}$ ratio obtained in our study are qualitatively consistent with the results in Refs.~\cite{Xu10,Xu13} based on a thermal model. 

It can also be seen in Fig.~\ref{Fig:Ratio} that the prediction on the $\pi^-/\pi^+$ ratio from the RVUU model with both the threshold effect and the pion in-medium effect is slightly larger than the upper value of  experimental data.  Using a softer symmetry energy with the slope parameter  $L=59~\mathrm{MeV}$, the $\pi^-/\pi^+$ ratio is reduced and becomes consistent with the experimental results. This result seems to contradict predictions by other transport models~\cite{Xia09, Fen10, Xie13} that a softer symmetry energy leads to a larger charged pion ratio. However, as indicated in Ref.~\cite{Son15}, the threshold effect reverses the effect of the symmetry energy on the $\pi^-/\pi^+$ ratio, namely, with the inclusion of the threshold effect, the softer the symmetry energy is, the smaller is the $\pi^-/\pi^+$ ratio. This is because the threshold effect is related to the density dependence of the symmetry energy, and becomes smaller with the softening of the symmetry energy, which then decreases the charged pion ratio. 

\begin{figure} [!hbt]
\centering 
\includegraphics[width=0.9\columnwidth]{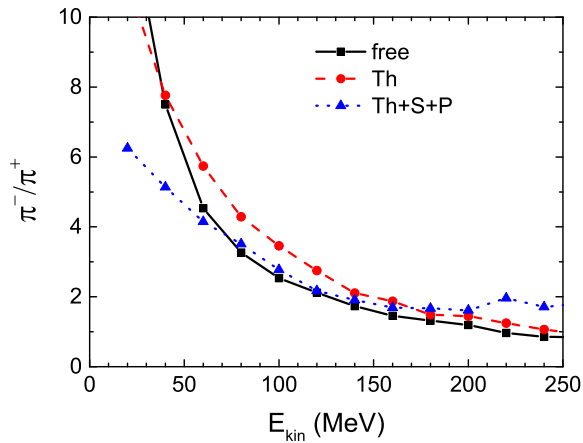} 
\caption{ (Color online) The $\pi^{-}/\pi^{+}$ ratio in Au + Au collisions at $E/A=400~\mathrm{MeV}$ and impact parameter of $1.4~\mathrm{fm}$ as a function of kinetic energy in the center-of-mass frame for the three cases of without any medium effects (free), with the in-medium threshold effect (Th), and with both the threshold and pion potential effects (Th+S+P). } \label{Fig:Spec}
\end{figure}

Recent studies have shown that the ratio of high-energy charged pions, i.e., their spectral ratio, is more sensitive to the high-density behavior of  nuclear symmetry energy~\cite{Hon14, Tsa16}. To study if this  remains the case after including the medium effects, we depict in Fig.~\ref{Fig:Spec} the $\pi^{-}/\pi^{+}$ ratio in Au + Au collisions at $E/A=400~\mathrm{MeV}$ and impact parameter of $1.4~\mathrm{fm}$ as a function of the kinetic energy $E_{\mathrm{kin}}$ of pion in the center-of-mass frame for the three cases of ``free'',``Th'', and ``Th+S+P''. It is seen that the threshold effect enhances  the charged pion ratio at energies $E_{\mathrm{kin}}>50~\mathrm{MeV}$. Including also pion potentials suppresses the $\pi^-/\pi^+$ ratio at energies $E_{\mathrm{kin}}<70~\mathrm{MeV}$ but enhances the ratio at higher energies. In particular, at $E_{\mathrm{kin}}=250~\mathrm{MeV}$, including both the threshold effect and pion potentials increases the $\pi^-/\pi^+$ ratio by a factor of 2, which is comparable to the effect due to the stiffness of  nuclear symmetry energy~\cite{Hon14, Tsa16}.

\section{Summary}

We have extended the relativistic Vlasov-Uehling-Uhlenbeck model based on the nonlinear relativistic NL$\rho$ mean-field model by including the isospin-dependent pion $s$-wave and $p$-wave potentials in nuclear medium, which are obtained from calculations based on the chiral perturbation theory and the $\Delta$-hole model, respectively. Their effects on the $\pi^-/\pi^+$ ratio  in Au + Au collisions at $E/A=400~\mathrm{MeV}$ have been studied. While the $\pi^-/\pi^+$ ratio is enhanced by the pion $p$-wave potential, it is significantly suppressed by the pion $s$-wave potential.  As a result, the pion potentials in nuclear medium lead to a significant reduction ($\sim10\%$) of  the $\pi^-/\pi^+$ ratio, which is comparable to that due to the stiffness of nuclear symmetry energy at high densities.  After including both the threshold effect and the pion in-medium effect, the $\pi^-/\pi^+$ ratio obtained from the RVUU model based on the relativistic NL$\rho$ model, which has a value of $L=84$ MeV for the slope parameter of nuclear symmetry energy, is slightly larger than the experimental upper value from the FOPI Collaboration. Using a softer symmetry energy of $L=59~\mathrm{MeV}$,  which is consistent with currently known empirical value~\cite{Oer16,BALi17}, by reducing the $\rho$-nucleon coupling constant in the NL$\rho$ model  can, however, well reproduce the experimental data on the charged pion ratio. 

We have also studied the effects of medium modification of thresholds and pion potentials on the $\pi^-/\pi^+$ spectral ratio. We have found that including these effects reduces the $\pi^-/\pi^+$ ratio at low pion kinetic energies, but considerably increases that at high pion kinetic energies as in Refs.~\cite{Hon14, Tsa16}. Our study thus indicates that the isospin-dependent in-medium threshold and pion potential effects need to be seriously treated in transport models to  describe more realistically pion production in heavy ion collisions, so that the behavior of nuclear symmetry energy at high density can be more reliably determined from the  charged pion ratio measured in these collisions.

\section*{Acknowledgements}

We thank Taesoo Song and Yifeng Sun for helpful communications and discussions. This work was supported by the US Department of Energy under Contract No. DE-SC0015266 and the Welch Foundation under Grant No. A-1358.

\end{document}